\documentclass[preprint,showpacs,preprintnumbers,amsmath,amssymb,nofootinbib]{revtex4}

\usepackage{graphicx}
\usepackage{dcolumn}
\usepackage{bm}
\usepackage{longtable}
\usepackage{latexsym}
\newcommand{\g}[1]{{\gamma}^{#1}}
\newcommand{\gc}[1]{{\gamma}_{#1}}

\newcommand{\diff}[2]{\frac{\delta {#1}}{\delta {#2}}}

\newcommand{\ket}[1]{|{#1}\rangle}
\newcommand{\bra}[1]{\langle {#1}|}
\newcommand{\inp}[2]{\langle{#1}\ket{#2}}
\newcommand{\vev}[1]{\langle{#1}\rangle}

\newcommand{\axialjc}[2]{j\sp{#1}\sb{A{#2}}}
\newcommand{\vectorjc}[2]{j\sp{#1}\sb{V{#2}}}

\newcommand{\smatrix}{\hat{\cal S}}

\newcommand{\fpi}{f_\pi}
\newcommand{\tfpi}{\tilde{f}_\pi}
\newcommand{\mpi}{m\sb{\pi}}

\newcommand{\ubexv}[2]{{\underline v}\sp{#1}\sb{#2}}

\newcommand{\ubexav}[2]{{\underline a}\sp{#1}\sb{#2}}
 
\newcommand{\ubexs}[1]{{\underline s}\sp{#1}}
\newcommand{\obexs}[1]{{\overline s}\sp{#1}}
\newcommand{\ubexp}[1]{{\underline p}\sp{#1}}
\newcommand{\obexp}[1]{{\overline p}\sp{#1}}
\newcommand{\ubexJ}[1]{{\underline J}\sp{#1}}

\newcommand{\uli}{\underline}
\newcommand{\oli}{\overline}

\begin{document}

\title{Master formula approach to broken chiral U(3)$\times$U(3) symmetry}

\author{Hiroyuki Kamano}
\affiliation{Thomas Jefferson National Accelerator Facility, Newport News, Virginia 23606, USA}


\begin{abstract}
The master formula approach to chiral symmetry breaking 
proposed by Yamagishi and Zahed is extended to 
the U$_R$(3)$\times$U$_L$(3) group in which effects of 
the U$_A$(1) anomaly and flavor symmetry breaking, $m_u\not=m_d\not=m_s$,
are taken into account.
New identities for the gluon topological susceptibility and
$\pi^0,\eta,\eta'\to\gamma^{(\ast)}\gamma^{(\ast)}$
decays are derived, which embody the consequences of
broken chiral symmetry in QCD without relying on any unphysical limits.
\end{abstract}



\maketitle

\section{Introduction}
\label{sec1}

Phenomena of the $\eta$ and $\eta^\prime$ mesons have been
an attractive subject both theoretically and experimentally.
Various properties of these mesons are closely related to 
the U$_A$(1) anomaly of QCD and the flavor mixing arising from
the mass differences among $u,d,s$ quarks.
Since the U$_A$(1) anomaly is a realization of the nontrivial topology 
of the gluonic configurations, investigating reactions and decaying 
processes associated with the $\eta$ and $\eta'$ mesons will be
one of the most feasible approaches to accessing the fundamental aspects of QCD.

Several theoretical approaches, such as
the chiral perturbation theory~\cite{Kai00,Bor01}
and the zero-momentum Ward identities with PCAC hypothesis~\cite{Sho00,Sho06},
have been applied for analyzing reaction processes
of $\eta$ and $\eta'$ mesons and exploring the gluonic context of the
low energy QCD. 
Most of these investigations start from certain unphysical limits
such as chiral, soft pion, and large $N_c$.

The master formula approach to chiral symmetry breaking proposed 
by Yamagishi and Zahed is a powerful tool
for analyzing hadronic processes 
which include ground state pseudoscalar mesons~\cite{Yam96,Lee99}.
The approach is based on a set of master equations, which fully takes 
account of the consequences of broken chiral symmetry without 
relying on any unphysical limits or expansion schemes.
Also, the master equations provide a systematic procedure, called 
the chiral reduction formula ($\chi$RF), to derive
the chiral Ward identities satisfied by scattering amplitudes
involving any number of pions with their physical masses.
The advantage of this approach is that
one can investigate the reaction dynamics, which cannot be determined
by chiral symmetry, separately
from the general structure required by the broken chiral
symmetry. 
Once such a separation is made, any models and/or expansion schemes 
can be employed for describing the reaction dynamics 
without any contradiction with the constraints from broken chiral symmetry.
A number of investigations based on the approach has been carried out 
for hadron reactions in the resonance 
region~\cite{Ste97,Ste98,Lee99,Kam04,Kam05-1,Kam06-1}
and hadronic matter~\cite{Ste96,Ste97-2,Lee98,Ste99,Dus07,Dus09,Dus09-2}.

Up to now, the approach has been formulated within the
two-flavor SU$_R$(2) $\times$ SU$_L$(2)~\cite{Yam96} and
three-flavor SU$_R$(3) $\times$ SU$_L$(3)~\cite{Lee99} groups 
in the isospin symmetric limit.
To analyze reaction processes involving $\eta'$
and address the features of the low energy QCD mentioned above, however,
we need to extend the approach to the U$_R$(3)$\times$U$_L$(3) group
incorporating the full flavor symmetry breaking due to 
$m_u\not=m_d\not=m_s$ and the U$_A$(1) anomaly.
In this paper we will describe how to make such an extension.

This paper is organized as follows.
In Sec.~\ref{sec2} we first review the master formula approach
for the isospin symmetric SU$_R$(2)$\times$SU$_L$(2) group
proposed in Ref.~\cite{Yam96}
and outline a derivation of the master equations with this simplest case.
Then in Sec.~\ref{sec3} we describe the extension to
the U$_R$(3)$\times$U$_L$(3) group including finite quark masses and the
U$_A$(1) anomaly.
Several applications of the U$_R$(3)$\times$U$_L$(3) master equations
are presented in Sec.~\ref{sec4}.
Summary and outlook are given in Sec.~\ref{sec5}.

\section{Review of isospin symmetric SU(2)$\times$SU(2) case}
\label{sec2}

\subsection{The Veltman-Bell equations}
\label{sec2a}

Consider QCD with massive $u$ and $d$ quarks 
in the isospin symmetric limit: $m_u =m_d=\hat m$.
The Lagrangian can be written as
\begin{equation}
{\cal L}^{SU(2)}_{\text{QCD}}=
[{\cal L}^{SU(2)}_{\text{QCD}}]_{(0)}
+\bar{q}\g{\mu}(v_\mu^a+\gc{5}a_\mu^a)\frac{\tau^a}{2}q(x)
-\bar{q}(\hat m + s - ip^a\gc{5}\tau^a) q(x),
\label{su2qcd}
\end{equation}
where $[{\cal L}^{SU(2)}_{\text{QCD}}]_{(0)}$
is the QCD Lagrangian in which the quark masses
are set to zero; $\tau^a$ ($a=1,2,3$) is the Pauli matrix for isospin; 
$q(x)$ is the isodoublet quark field $q = (u,d)^T$.
The vector, axial-vector, scalar,
and pseudoscalar external fields, $\phi = (v_\mu^a,a_\mu^a,s,p^a)$,
are treated as sources to generate 
the corresponding currents and densities,
${\cal O} = (V^a_\mu,A^a_\mu,\Sigma,\Pi^a)$,
which can be defined by 
${\cal O}=\delta(\int d^4x{\cal L}^{SU(2)}_{QCD})/\delta\phi$.

A fundamental quantity in the theoretical framework 
developed in Ref.~\cite{Yam96}
is the extended S-matrix operator ${\cal S}[\phi]$,
a functional of the external fields $\phi$. 
This operator is unitary, ${\cal S}^\dag {\cal S}={\cal S}{\cal S}^\dag=1$,
and is related to the vacuum-to-vacuum transition amplitude 
in the presence of the external fields $\phi$:
$Z[\phi] = \inp{0\text{ out}}{0\text{ in}}_{\phi}
= \bra{0\text{ in}}{\cal S}[\phi]\ket{0\text{ in}}
= \bra{0\text{ out}}{\cal S}[\phi]\ket{0\text{ out}}$.
The Schwinger action principle allows one to express the 
quantum operators corresponding to ${\cal O}=(V^a_\mu,A^a_\mu,\Sigma,\Pi^a)$
as~\cite{Yam96,Yam89,Bog}
\begin{equation}
{\cal O}(x) = -i{\cal S}^\dag\diff{}{\phi(x)}{\cal S},
\label{operator}
\end{equation}
and their $T^\ast$-product as
\begin{equation}
T^\ast[{\cal O}(x_1)\cdots{\cal O}(x_n)] =
(-i)^n{\cal S}^\dag \diff{}{\phi(x_1)}\cdots\diff{}{\phi(x_n)} {\cal S}.
\label{tproduct}
\end{equation}
A more detailed description of
the theoretical formulation based on the extended S-matrix
can be found in the literature~\cite{Bog,Yam89,Yam87}
and this will not be discussed here.

The system described by Eq.~(\ref{su2qcd}) and its effective theory
have an approximate SU$_R$(2)$\times$SU$_L$(2) 
chiral symmetry explicitly broken by the quark masses. 
It is known that such systems satisfy the divergence equations for 
the vector and axial currents, which, 
following Ref.~\cite{Yam96},
we call the Veltman-Bell (VB) equations~\cite{Vel66,Bel67}. 
The explicit forms are
\begin{equation}
\nabla^{\mu ac}V_\mu^c + \ubexav{\mu ac}{}A_\mu^c + \ubexp{ac}\Pi^c =0,
\label{su2vvb}
\end{equation}
\begin{equation}
\nabla^{\mu ac}A_\mu^c + \ubexav{\mu ac}{}V_\mu^c 
- (\hat m + s)\Pi^a + p^a\Sigma =0,
\label{su2avb}
\end{equation}
where we have introduced the notation
$\underline{X}^{ac}=\varepsilon^{abc}X^b$ (applicable
to any quantity with one-isospin index $X^b$);
$\nabla^{ac}_\mu = \delta^{ac}\partial_\mu + \ubexv{ac}{\mu}$.

With Eq.~(\ref{operator}),
the VB equations~(\ref{su2vvb}) and (\ref{su2avb})
can be rewritten as a set of linear equations of
the extended S-matrix:
\begin{equation}
\left[
\nabla^{ac}_\mu\diff{}{v^c_\mu}
+\ubexav{ac}{\mu}\diff{}{a^c_\mu}
+\ubexp{ac}\diff{}{p^c}
\right]
{\cal S} =0,
\label{su2xvs}
\end{equation}
\begin{equation}
\left[
\nabla^{ac}_\mu\diff{}{a^c_\mu}
+\ubexav{ac}{\mu}\diff{}{v^c_\mu}
-(s+\hat m)\diff{}{p^a}
+p^a\diff{}{s} 
\right]
{\cal S} =0.
\label{su2xas}
\end{equation}
Applying functional derivatives of $\phi$ to 
Eqs.~(\ref{su2xvs}) and (\ref{su2xas})
and using Eqs.~(\ref{operator}) and (\ref{tproduct}),
one can derive the vector and axial Ward identities
satisfied by the operators ${\cal O} = (V^a_\mu,A^a_\mu,\Sigma,\Pi^a)$.

\subsection{Master equations for the chiral symmetry breaking}
\label{sec2b}

The VB equations~(\ref{su2vvb}) and~(\ref{su2avb}) 
[or equivalently Eqs.~(\ref{su2xvs}) and~(\ref{su2xas})]
are satisfied by the systems
in both the Wigner and Nambu-Goldstone (NG) phases.
However, if the chiral symmetry in the system
is spontaneously broken, the NG bosons (pions) appear and 
couple to the axial current and the isovector-pseudoscalar density:
\begin{equation}
\bra{0} A_\mu^a(x)\ket{\pi^b(p)} = \fpi ip_\mu \delta^{ab}e\sp{-ipx},
\label{anrmsu2}
\end{equation}
\begin{equation}
\bra{0} \Pi^a(x)\ket{\pi^b(p)} = G\delta^{ab}e\sp{-ipx}.
\label{pnrmsu2}
\end{equation}
Here $\fpi$ is the pion decay constant, which remains finite in the chiral
limit in contrast to those of the pion excitations~\cite{Hol04};
$G$ is the pseudoscalar coupling constant. 
By taking the matrix element of the axial VB equations~(\ref{su2avb})
between the vacuum state $\bra{0}$ and the one-pion state 
$\ket{\pi^a(p)}$, and using Eqs.~(\ref{anrmsu2}) and~(\ref{pnrmsu2}),
we obtain the following mass relation:
\begin{equation}
\fpi \mpi^2 = \hat m G,
\label{mass-rel-su2}
\end{equation}
which shows clearly the well-known result that the nonzero quark masses
are responsible for the nonzero pion masses.
The master equations proposed in Ref.~\cite{Yam96} may be understood
as the VB equations incorporating this information on
chiral symmetry breaking.

It was shown in Ref.~\cite{Yam96} that
the information can be incorporated into
the VB equations by making the following modifications:
\begin{enumerate}
\item
Introduce new pseudoscalar and scalar external fields,
$J^a$ and $Y$, defined by
\begin{equation}
J^a= Gp^a + \fpi\nabla^{ac}_\mu a^{\mu c},
\label{su2-mod11}
\end{equation}
\begin{equation}
Y= Gs,
\label{su2-mod12}
\end{equation} 
and treat $\phi = (a^a_\mu,v^a_\mu,Y,J^a)$ as independent external fields.
\item Introduce a new extended S-matrix $\smatrix$ as
\begin{equation}
\smatrix = {\cal S} \exp (-i\delta {\cal I}),
\label{su2-mod21}
\end{equation}
with
\begin{equation}
\delta{\cal I}=
\int d^4x\left[Y(x)G^{-1}C+\frac{\fpi^2}{2}a^{\mu a}(x)a^a_\mu(x)\right],
\label{su2-mod22}
\end{equation}
where a new constant $C$ is introduced. 
(The physical meaning of $C$ is explained below.)
\end{enumerate}
With these modifications, the new current and density operators,
$\hat{\cal O} = (\vectorjc{a}{\mu},\axialjc{a}{\mu},
\hat{\sigma},\hat{\pi}\sp{a})$,
defined by 
$\hat{\cal O}=-i\smatrix^\dag(\delta/\delta\phi)\smatrix$,
are related with the original current and density operators,
${\cal O}=(V^a_\mu,A^a_\mu,\Sigma,\Pi^a)$, as follows:
\begin{equation}
V^a_\mu =  \vectorjc{a}{\mu} + \fpi \ubexav{ac}{\mu}\hat\pi^c,
\label{su2-newv}
\end{equation}
\begin{equation}
A^a_\mu =  \axialjc{a}{\mu}+\fpi^2 a^a_\mu - \fpi\nabla^{ac}_\mu\hat\pi^c,
\label{su2-newa}
\end{equation}
\begin{equation}
\Sigma  =  G\hat\sigma + C,
\label{su2-news}
\end{equation}
\begin{equation}
\Pi^a = G\hat\pi^a.
\label{su2-newp}
\end{equation}
Here the new pseudoscalar density $\hat\pi^a$ satisfies
$\bra{0}\hat\pi^a(x)\ket{\pi^b(p)}=\delta^{ab}e^{-ipx}$.
This allows one to identify $\hat\pi^a$ with the normalized
interpolating pion field.
The change of the field variable $p^a\to J^a$ defined by Eq.~(\ref{su2-mod11}) 
is responsible for the separation of the one-pion component from
the axial current $A_\mu^a$ and thus is introduced to
indicate that the system under consideration includes pions.
The new axial current $\axialjc{a}{\mu}$ has
no one-pion component surviving on pion mass-shell,
$\bra{0}\axialjc{a}{\mu}(x)\ket{\pi^b(p)}=0$,
in contrast to the original $A_\mu^a$.
 
The second modification defined by 
Eqs.~(\ref{su2-mod21}) and~(\ref{su2-mod22}) is introduced to
indicate that the action describing the original S-matrix operator, 
${\cal S}$, should contain contact terms, as defined in Eq.~(\ref{su2-mod22}),
if the chiral symmetry is spontaneously broken.
A justification of this statement has been provided
in Ref.~\cite{Yam96} making use of the gauged nonlinear sigma model.
It is also noted that the difference between ${\cal S}$ and $\hat{\cal S}$
just comes from the contact terms of the $c$-number external fields and thus
does not affect the physical observables.

The practical role of the second modification is twofold.
The first term in the integrand of $\delta {\cal I}$
is introduced to take account of the quark-antiquark condensation.
In fact, it explicitly introduces a shift in the scalar
density [see Eq.~(\ref{su2-news})], amounting to the new constant, $C$. 
This constant carries part of the information on
the condensation. (We allow $\vev{\hat\sigma}\not=0$.)
It turns out that $C$ is expressed as the product of the pion decay constant 
and the pseudoscalar coupling constant:
\begin{equation}
C=\fpi G.
\label{su2-cfg}
\end{equation}
This follows from the fact that $\hat\pi^a$ is the normalized
interpolating pion field.
(See Appendix~\ref{app1} for the derivation.)
The mass relation~(\ref{mass-rel-su2}) can be rewritten as
\begin{equation}
\fpi^2\mpi^2 = \hat m C.
\label{mass-rel-su2-2}
\end{equation}
If $\vev{\hat\sigma}=0$, then $C=\vev{\Sigma}$ 
and Eq.~(\ref{mass-rel-su2-2}) 
reduces to the Gell-Mann-Oakes-Renner (GMOR) relation.
Therefore, $\vev{\hat\sigma}$ represents the deviation
of the mass relation from the GMOR relation.
In fact, the on-shell ($1/\fpi$) expansion scheme proposed in 
Ref.~\cite{Yam96}, which is the expansion in $1/\fpi$ 
around the physical pion mass
and is constructed so that the GMOR relation holds at the leading order,
leads to $\vev{\hat\sigma}=0+{\cal O}(\fpi^{-1})$
and $C=\vev{\Sigma}+{\cal O}(\fpi^{-1})$.
The second term in the integrand of $\delta {\cal I}$
results in the appearance of $\fpi^2 a_\mu^a$ in Eq.~(\ref{su2-newa}).
This ensures the existence of the contact term
$\delta^{ab}\fpi^2 g_{\mu\nu}$ in the two-point function
of the axial current,
\begin{equation}
i\int d^4x e^{iqx}\vev{T^\ast[A_\mu^a(x)A_\nu^b(0)]} =
\delta^{ab}\fpi^2g_{\mu\nu} -\delta^{ab}\fpi^2\frac{q_\mu q_\nu}{q^2-\mpi^2}
+\cdots,
\label{su2-AA}
\end{equation}
which must appear in the two-point function to have a correct chiral limit.
(The symbol $\vev{~~}$ denotes the vacuum expectation value.)

Substituting Eqs.~(\ref{su2-mod11}),~(\ref{su2-mod12}), 
and~(\ref{su2-newv})-(\ref{su2-newp}) into
the VB equations and using Eqs.~(\ref{mass-rel-su2})
and~(\ref{su2-cfg}), we have
\begin{equation}
[\nabla^{\mu ac}\vectorjc{c}{\mu}
+\ubexav{\mu ac}{}\axialjc{c}{\mu}+\ubexJ{ac}{}\hat\pi^c]= 0,
\end{equation}
\begin{eqnarray}
\left[
- \nabla^{\mu ae}\nabla_\mu^{ec} + \ubexav{\mu ae}{}\ubexav{ec}{\mu}
- \mpi^2\delta^{ac}
- Y\fpi^{-1}\delta^{ac}
\right]
\hat{\pi}\sp{c}
&=&
\nonumber\\
&&
\!\!\!\!\!\!\!\!\!\!\!\!\!\!\!\!\!\!\!\!
\!\!\!\!\!\!\!\!\!\!\!\!\!\!\!\!\!\!\!\!
\!\!\!\!\!\!\!\!\!\!\!\!\!\!\!\!\!\!\!\!
-J^a(x)
-\fpi^{-1}(\nabla^{\mu ac}\axialjc{c}{\mu}
+\ubexav{\mu ac}{}\vectorjc{c}{\mu})
-[J^a - \fpi\nabla^{\mu ac}a^c_\mu] \hat\sigma.
\label{su2pioneq}
\end{eqnarray}
These equations can be written 
in the functional derivative form as
\begin{equation}
\left(
\nabla^{ac}_\mu\diff{}{v^c_\mu}+\ubexav{ac}{\mu}\diff{}{a^c_\mu}
+\ubexJ{ac}\diff{}{J^c}
\right)
\smatrix = 0,
\label{su2tvs}
\end{equation}
\begin{equation}
\left[
-( \Box\delta^{ab} + \mpi^2\delta^{ab} + K_{\text{SU(2)}}^{ab} ) \diff{}{J^b}
+ iJ^a
+\fpi^{-1}t^a_{A}
-\left(\nabla^{\mu ac}a^c_\mu-\fpi^{-1}J^a\right)\diff{}{Y}
\right]\smatrix=0,
\label{su2tas}
\end{equation}
with
\begin{equation*}
K_{\text{SU(2)}}^{ab} =
\nabla^{\mu ac}\nabla^{cb}_\mu - \ubexav{ac}{\mu}\ubexav{\mu cb}{} 
+ \delta^{ab}Y- \delta^{ab}\Box,
\ \ \  
t_A^a =
 \nabla^{ac}_\mu\diff{}{a^c_\mu}
+\ubexav{ac}{\mu}\diff{}{v^c_\mu}.
\end{equation*}
By introducing the retarded and advanced Green functions satisfying
\begin{equation*}
-[\Box\delta^{ab} + \mpi\sp{2}\delta^{ab} + K_{\text{SU(2)}}^{ab}(x)]
G\sp{bc}\sb{R,A}(x,y)
=\delta\sp{ac}\delta\sp{(4)}(x-y),
\end{equation*}
the axial VB equation~(\ref{su2pioneq}) can be formally solved for
the interpolating pion field $\hat\pi$.
The functional derivative form of the solution is written as
\begin{eqnarray}
\diff{\smatrix}{J^a(x)} &=& 
 i \smatrix \pi_{\text{in}}^a(x)
+i \smatrix \int d^4y
   G_R^{ab}(x,y)K_{\text{SU(2)}}^{bc}(y)\pi^c_{\text{in}}(y)
- \int d^4y G_R^{ab}(x,y)\bar R^b_{\text{SU(2)}}(y) \smatrix
\nonumber\\
&=&
 i \pi_{\text{in}}^a(x)\smatrix
+i \int d^4y
   G_A^{ab}(x,y)K_{\text{SU(2)}}^{bc}(y)\pi^c_{\text{in}}(y)\smatrix
- \int d^4y G_A^{ab}(x,y)\bar R^b_{\text{SU(2)}}(y) \smatrix.
\label{su2js}  
\end{eqnarray}
Here $\pi^a_{\text{in}}$ is the in-state asymptotic pion field,
$\hat\pi^a \to \pi^a_{\text{in}}+\cdots(t\to -\infty)$;
$\bar R^a_{\text{SU(2)}}(x)$ is defined by
\begin{equation}
\bar R^a_{\text{SU(2)}}(x)= 
R_{\text{SU(2)}}^a(x) +K_{\text{SU(2)}}^{ab}(x)\diff{}{J^b(x)},
\label{su2rbar}
\end{equation}
with
\begin{equation}
R_{\text{SU(2)}}^a(x) =
\left[
 iJ^a
+ \frac{1}{\fpi} t^a\sb{A}
- K_{\text{SU(2)}}^{ab}\diff{}{J^b} 
- \left( \nabla^{\mu ac}a_\mu^{c} - \frac{J^a}{\fpi} \right) \diff{}{Y}
\right](x).
\label{su2Rx}
\end{equation}
Equations~(\ref{su2tvs}) and~(\ref{su2js}) constitute the master equations
for the SU$_R$(2)$\times$SU$_L$(2) chiral symmetry breaking 
in the isospin symmetric limit.

\subsection{Chiral reduction formula}
\label{sec2d}

From the axial master equation~(\ref{su2js}), we can derive 
the commutation relations between the creation and annihilation operators
of the pion and the extended S-matrix $\smatrix$,
\begin{equation}
[ a\sp{a}\sb{{\rm in}}(k), \smatrix ] = R_{\text{SU(2)}}^a(k)\smatrix, 
\ \ \ \  
[ \smatrix , a\sp{a\dag}\sb{{\rm in}}(k) ] = R_{\text{SU(2)}}\sp{a}(-k)\smatrix,
\label{su2comm}
\end{equation}
where $R_{\text{SU(2)}}^a(k) =\int d^4x e^{ikx}R_{\text{SU(2)}}^a(x)$.
Note that we can rewrite the commutation relations
in this form without using the asymptotic pion field.
[Compare them with Eqs.~(6.2) and (6.3) in Ref.~\cite{Yam96}.]

Iterative use of Eq.~(\ref{su2comm}) results 
in the $\chi$RF for the 
on-shell scattering amplitudes 
involving any number of pions with their physical masses,
\begin{eqnarray}
\bra{\alpha;k\sb{1}a\sb{1},\cdots,k\sb{m}a\sb{m}} 
\smatrix
\ket{\beta;l\sb{1}b\sb{1},\cdots,l\sb{n}b\sb{n}}|\sb{\phi=0}
&=&
\nonumber\\
&&
\!\!\!\!\!\!\!\!\!\!\!\!\!\!\!
\!\!\!\!\!\!\!\!\!\!\!\!\!\!\!
\!\!\!\!\!\!\!\!\!\!\!\!\!\!\!
\!\!\!\!\!\!\!\!\!\!\!\!\!\!\!
\!\!\!\!\!\!\!\!\!\!\!\!\!\!\!
\!\!\!\!\!\!\!\!\!\!\!\!\!\!\!
[ R_{\text{SU(2)}}\sp{a\sb{1}}(k\sb{1})  \cdots R_{\text{SU(2)}}\sp{a\sb{m}} (k\sb{m}) 
  R_{\text{SU(2)}}\sp{b\sb{1}}(-l\sb{1}) \cdots R_{\text{SU(2)}}\sp{b\sb{n}} (-l\sb{n}) ]\sb{{\rm S}} 
\bra{\alpha} \smatrix \ket{\beta}|\sb{\phi=0},
\label{su2chrf}
\end{eqnarray}
where $k\sb{i}~(l\sb{i})$ and $a\sb{i}~(b\sb{i})$ are, respectively,
the four momentum and isospin indices of the outgoing (incoming) pions; 
$\alpha$ and $\beta$ stand for states of other particles.
Here we consider the case that no two pions have equal momenta.
The symbol $[~~]\sb{{\rm S}}$ represents normalized symmetric permutations
of the functional derivative operators,
\begin{equation}
[{\cal D}\sb{1}\cdots{\cal D}\sb{n}]\sb{{\rm S}} =
\frac{1}{n!} \sum\sb{{\rm perms.}} {\cal D}\sb{1}\cdots{\cal D}\sb{n}.
\label{sym-perm}
\end{equation}
This operation shows clearly the crossing symmetry in Eq.~(\ref{su2chrf}).
By using Eq.~(\ref{su2chrf}) together with Eq.~(\ref{tproduct}),
scattering amplitudes are expressed in terms of Green's functions of 
the operators 
$\hat{\cal O}=(\vectorjc{a}{\mu},\axialjc{a}{\mu},\hat{\sigma},\hat{\pi}^a)$.
The $\chi$RF takes the form of functional derivatives, 
and all constraints which stem from broken chiral symmetry 
are contained in $R^a_{\text{SU(2)}}(k)$.
The extension of the $\chi$RF to 
the off-shell pions has been discussed in detail in Ref.~\cite{Kam06-2}.

\section{Extending to U$_R$(3)$\times$U$_L$(3) 
with flavor symmetry breaking}
\label{sec3}

In this section, we extend the master equations
reviewed in Sec.~\ref{sec2} to the U$_R$(3)$\times$U$_L$(3) group
with $m_u\not=m_d\not=m_s$.
In the remainder of this paper, the term ``pion'' (``$\pi$'') is used
for expressing the nonet ground pseudoscalar mesons generically,
and the symbols $\pi^{\pm,0}$, $K^{+,0}$, $\bar{K}^{-,0}$, $\eta$, and $\eta'$
are used for referring to the specific mesons.

\subsection{The U$_R$(3)$\times$U$_L$(3) VB equations}
\label{sec3a}

The Lagrangian now includes massive $u, d, s$ quarks:
\begin{equation}
{\cal L}_{\text{QCD}}^{U(3)}
=
[{\cal L}_{\text{QCD}}^{U(3)}]_{(0)}
+\bar{q}\g{\mu}(v_\mu^a+\gc{5}a_\mu^a)\frac{\lambda^a}{2}q
-\bar{q}(m_q^a + s^a - ip^a\gc{5})\lambda^a q-\theta\omega.
\label{u3qcd}
\end{equation}
Here $q^T=(u,d,s)$; $m_q^a\lambda^a = \text{diag}(m_u,m_d,m_s)$;
$\theta$ and $\omega$ are the vacuum angle and
the gluon topological charge density, respectively. 
The term $[{\cal L}_{\text{QCD}}^{U(3)}]_{(0)}$ represents
the QCD Lagrangian with the quark masses, the vacuum angle, 
and all external fields set to zero.
The flavor matrix $\lambda^a$ is taken to be one of the Gell-Mann matrices 
for $a=1,\cdots,8$ and $\lambda^0 = \sqrt{2/3}{\bf 1}$
so that they satisfy $\text{Tr}[\lambda^a\lambda^b]=2\delta^{ab}$.
The vector, axial-vector, scalar, and pseudoscalar external
fields and the vacuum angle,
$\phi = (v_\mu^a,a_\mu^a,s^a,p^a,-\theta)$,
are sources to generate currents and densities,
${\cal O} = (V_\mu^a,A_\mu^a,\Sigma^a,\Pi^a,\omega)$,
which are obtained from
${\cal O} = \delta(\int d^4x{\cal L}^{U(3)}_{\text{QCD}})/\delta\phi$.
As in the SU(2) case,
the quantum operators corresponding to ${\cal O}$ are defined by 
Eq.~(\ref{operator}).

The VB equations for the U$_R$(3)$\times$U$_L$(3) group can then 
be written as
\begin{equation}
\nabla^{\mu ac}V^c_\mu
+ \ubexav{\mu ac}{} A^c_\mu
+ \ubexp{ac} \Pi^c
+ (\ubexs{} + \underline{m}_q)^{ac} \Sigma^c
=0,
\label{u3vvb}
\end{equation}
\begin{equation}
\nabla^{\mu ac}A^c_\mu
+ \ubexav{ab}{\mu} V^b_\mu
+ \obexp{ab}\Sigma^b
- (\obexs{} + \overline{m}_q)^{ac}\Pi^c
- \text{Tr}[\lambda^a]\omega
= \Omega^a.
\label{u3avb}
\end{equation}
Here we have introduced the notation\footnote{
The structure constants $f^{abc}$ and $d^{abc}$ are defined by
$f^{abc}=-(i/4)\text{Tr}[[\lambda^a,\lambda^b]\lambda^c]$ and
$d^{abc}=(1/4)\text{Tr}[\{\lambda^a,\lambda^b\}\lambda^c]$, respectively.
With this definition, we have $f^{0bc} = 0$ and $d^{0bc}=\sqrt{2/3}\delta^{bc}$.
}
$\oli{X}^{ac}= d^{abc}X^b$ and
$\uli{X}^{ac}= f^{abc}X^b$ (applicable to any quantity
with one-flavor index $X^b$);
$\nabla^{ac}_\mu = \partial_\mu\delta^{ac} +\ubexv{ac}{\mu}$.
It is well known that an additional nonconserving 
contribution in the axial VB equation, which amounts to 
$- \text{Tr}[\lambda^a]\omega$, is attributable to the U$_A$(1) anomaly.
Furthermore, for the U$_R$(3)$\times$U$_L$(3) group,
the non-Abelian anomaly $\Omega^a$ associated 
with the external gauge fields $v_\mu^a$ and $a_\mu^a$ also appears.
Its explicit form is~\cite{Bar69}
\begin{eqnarray}
\Omega^a &=& 
\frac{N_c}{16\pi^2}\varepsilon^{\mu\nu\rho\sigma}
\text{Tr}
\left[
\frac{\lambda^a}{2}
\left(
  F^V_{\mu\nu} F^V_{\rho\sigma}
+ \frac{1}{3}F^A_{\mu\nu} F^A_{\rho\sigma}
\right.
\right.
\nonumber\\
&&
\left.
\left.
+ i\frac{8}{3}
(a_\mu a_\nu F^V_{\rho\sigma}
+a_\mu F^V_{\nu\rho} a_\sigma
+F^V_{\mu\nu} a_\rho a_\sigma)
-\frac{32}{3}a_\mu a_\nu a_\rho a_\sigma
\right)
\right],
\label{nonabelian}
\end{eqnarray}
with $F^V_{\mu\nu} = 
\partial_\mu v_\nu -\partial_\nu v_\mu -i [v_\mu,v_\nu]-i[a_\mu,a_\nu]$,
$F^A_{\mu\nu} = 
\partial_\mu a_\nu -\partial_\nu a_\mu -i [v_\mu,a_\nu]-i[a_\mu,v_\nu]$,
$v_\mu = v^a_\mu (\lambda^a/2)$, and
$a_\mu = a^a_\mu (\lambda^a/2)$.
We take a convention with $\varepsilon_{0123} = +1$.

The functional derivative form of the VB equations is given by
\begin{equation}
\left[\nabla_\mu^{ac}\diff{}{v_\mu^c}
+\ubexav{ac}{\mu}\diff{}{a_\mu^c}
+\ubexp{ac}\diff{}{p^c}
+(\ubexs{}+\uli{m}_q)^{ac}\diff{}{s^c}
\right]
{\cal S}=0,
\label{u3vvb2}
\end{equation}

\begin{equation}
\left[
\nabla_\mu^{ac}\diff{}{a_\mu^c}
+\ubexav{ac}{\mu}\diff{}{v_\mu^c}
+\obexp{ac}\diff{}{s^c}
-(\obexs{}+\oli{m}_q)^{ac}\diff{}{p^c}
+\text{Tr}[\lambda^a]\diff{}{\theta}
\right]
{\cal S} = i\Omega^a{\cal S}.
\label{u3avb2}
\end{equation}

\subsection{Flavor symmetry breaking and U$_A$(1) anomaly}
\label{sec3b}

Besides their key role in the singlet sector,
the mass differences among the $u,d,s$ quarks 
and the U$_A$(1) anomaly are responsible for the major
complication in the construction of the U$_R$(3)$\times$U$_L$(3)
master equations.
Therefore we first examine those effects carefully.

The mass differences of the quarks violate the flavor symmetry explicitly
as well as the chiral symmetry.
Thus the flavor U(3) basis cannot be the mass eigenstates of hadrons
and mixing of the U(3) basis occurs.
Necessary information on the flavor symmetry breaking
here is summarized in the following matrix elements,
\begin{equation}
\bra{0} A_\mu^a (x)\ket{P(p)} = \tfpi^{aP} ip_\mu e^{-ipx},
\label{anrm}
\end{equation}
\begin{equation}
\bra{0} \Pi^a (x)\ket{P(p)} = G^{aP} e^{-ipx},
\label{pnrm}
\end{equation}
\begin{equation}
\bra{0}\omega(x)\ket{P(p)}= \tilde{\cal A}^P e^{-ipx},
\label{onrm}
\end{equation}
where $P$ denotes the physical pion states (the mass eigenstates):
$P=(\pi^{\pm,0},K^{\pm,0},\bar K^0,\eta,\eta^\prime)$.
(Capital indices are used to represent the mass eigenstates.)
Because of the flavor symmetry breaking,
the pion decay constant $\tfpi^{aP}$ and 
the pseudoscalar coupling constant $G^{aP}$
now have two indices that specify
the flavor U(3) basis and the mass eigenstates\footnote{
The breaking pattern of the flavor symmetry, i.e.,
$m_q^a\lambda^a = m_q^0\lambda^0 +m_q^3\lambda^3+m_q^8\lambda^8$,
implies that $\tfpi^{aP}$ and $G^{aP}$ are block diagonal in 
$(1,2)\times(\pi^+,\pi^-)$,
$(4,5)\times(K^+,K^-)$,
$(6,7)\times(K^0,\bar K^0)$,
and $(0,3,8)\times(\pi^0,\eta,\eta')$.
In the isospin symmetric limit they become
$(1,2,3)\times(\pi^+,\pi^-,\pi^0)$,
$(4,5,6,7)\times(K^+,K^-,K^0,\bar K^0)$,
and $(0,8)\times(\eta,\eta')$ and the first two blocks can be diagonalized.
}.
Also, it is noted that $\tilde{\cal A}^P$ has a nonzero value 
only for $P=\pi^0,\eta,\eta'$.

From the axial VB equation~(\ref{u3avb}) and Eqs.~(\ref{anrm})-(\ref{onrm}),
we can derive the U(3) version of the mass relation:
\begin{equation}
\tfpi^{aQ}(\mpi^2)^{QP}=\oli{m}_q^{ab}G^{bP}+\text{Tr}[\lambda^a]\tilde{\cal A}^P,
\label{u3mass1}
\end{equation}
where $\mpi^2$ is the diagonal mass-squared matrix of the physical nonet pions. 
The appearance of $\tilde{\cal A}^P$ is attributable to $\omega$ 
in the axial VB equation and thus is nothing but 
a consequence of the U$_A$(1) anomaly. 

As already mentioned above, the U$_A$(1) anomaly leads to
the nonconserving term proportional to $\omega$ 
in the singlet axial VB equation.
Because of this, the operators $A_\mu^0$ and $\omega$
get renormalized under a change of the QCD scale~\cite{Kai00}:
$(A_\mu^0)_{\text{ren.}} = Z_A A_\mu^0$,
$(\omega)_{\text{ren.}} 
= \omega + (Z_A-1)(1/\text{Tr}[\lambda^0])\partial^\mu A_\mu^0$,
where $Z_A$ is the renormalization factor.
Accordingly,
the singlet pion decay constant $\tfpi^{0P}$ and $\tilde{\cal A}^P$
also get renormalized as
\begin{eqnarray}
(\tfpi^{0P})_{\text{ren.}}&=& Z_A \tfpi^{0P},
\label{fpiren}
\\
(\tilde{\cal A}^P)_{\text{ren.}}&=&
\tilde{\cal A}^P 
+ (Z_A - 1)\frac{1}{\text{Tr}[\lambda^0]}\tfpi^{0Q}(\mpi^2)^{QP}.
\label{calAren}
\end{eqnarray}
These are thus scale dependent and not physical constants.
This is in contrast to the octet pion decay constants, $\tfpi^{aP}$, 
with $a\not =0$, which are scale independent.
It is noted, however, that the combination 
$\tfpi^{0Q}(\mpi^2)^{QP}- \text{Tr}[\lambda^0]\tilde{\cal A}^P$ is 
invariant under a change of the QCD scale.

As pointed out by Shore and Veneziano (see, e.g., 
Refs.~\cite{Sho92-1,Sho06,Sho07}),
the renormalization group (RG) variant $\tfpi^{0P}$,
which is defined as 
a coupling of the singlet axial current to the physical pions,
does not satisfy GMOR type mass relations.
They introduced a RG-invariant decay constant
$\fpi^{0P}$ and showed that the use of this new constant
provides a natural extension of the GMOR relation 
to the singlet sector.

Following Shore and Veneziano, 
let us introduce the RG invariant $\fpi^{0P}$. 
Defining $\delta f_\pi^{0P}= \tfpi^{0P} - \fpi^{0P}$, 
the combination $(\tfpi\mpi^2)^{0P} - \text{Tr}[\lambda^0]\tilde{\cal A}^P$ 
can be rewritten as
\begin{equation}
(\tfpi\mpi^2)^{0P} - \text{Tr}[\lambda^0]\tilde{\cal A}^P =
(\fpi\mpi^2)^{0P} - \text{Tr}[\lambda^0][\tilde{\cal A}^P 
- (\delta\fpi\mpi^2)^{0P}].
\label{eq54}
\end{equation}
Because $(\tfpi\mpi^2)^{0P} - \text{Tr}[\lambda^0]\tilde{\cal A}^P$  and
$(\fpi\mpi^2)^{0P}$ are individually RG invariant,
${\cal A}^P\equiv [\tilde{\cal A}^P - (\delta\fpi\mpi^2)^{0P}]$
is also RG invariant.
The values of $\fpi^{0P}$ and ${\cal A}^P$ must be extracted from
the experimental data. 
With this modification, the mass relation~(\ref{u3mass1}) becomes
\begin{eqnarray}
(\fpi\mpi^2\fpi^T)^{ab} &=&
(\oli{m}_qG\fpi^T)^{ab}
+\text{Tr}[\lambda^a]({\cal A}\fpi^T)^b
\nonumber\\
&=&
(\oli{m}_qG\fpi^T)^{ab} +A^{ab},
\label{u3mass2}
\end{eqnarray}
where $\fpi^{aP}\equiv\tfpi^{aP}$ for $a\not = 0$;  
$(\fpi^T)=(\fpi^T)^{Pa}$ is the transpose of $\fpi^{aP}$;
$A^{ab}\equiv\text{Tr}[\lambda^a]({\cal A}\fpi^T)^b$.

\subsection{Master equations for the U$_R$(3)$\times$U$_L$(3) symmetry breaking}
\label{sec3d}

We can incorporate the information on the symmetry breaking
described in Sec.~\ref{sec3b} in a completely parallel way to the SU(2) case:
\begin{enumerate}
\item 
Introduce a new pseudoscalar and scalar external fields 
$J^P$ and $Y^P$ defined by
\begin{equation}
J^P  =
p^a G^{aP}
+ (\tfpi^T)^{Pa}(\nabla^\mu\tilde a_\mu)^a
-
\left[
(\fpi^T)^{P0}\Box + (\fpi^{-1})^{Pa} A^{a0}
\right] \frac{1}{\text{Tr}[\lambda^0]}\theta,
\label{J}
\end{equation}
\begin{equation}
Y^P= s^aG^{aP},
\label{Y}
\end{equation}
with 
$\tilde a^a_\mu = a^a_\mu +\delta^{a0}(1/\text{Tr}[\lambda^0])
\partial_\mu \theta$, 
and then consider $\phi=(a^a_\mu,v^a_\mu,Y^P,J^P,-\theta)$
as independent external field variables.
\item 
Introduce new extended S-matrix by
\begin{eqnarray}
\smatrix &=&{\cal S}\exp
\left[
-i \left(\delta{\cal I} + \delta{\cal I}_\Omega\right)
\right],
\label{u3smatrix}
\end{eqnarray}
where
\begin{eqnarray}
\delta {\cal I} &=&
 \int d^4x\left( Y^P(G^{-1})^{Pa}C^a
+ \frac{1}{2}\tilde a^{\mu a}(\tfpi\tfpi^T)^{ab}\tilde a_\mu^b
\right.
\nonumber\\
&&
\left.
- \frac{1}{2(\text{Tr}[\lambda^0])^2}\theta (A+\fpi\fpi^T \Box)^{00}\theta
-\frac{1}{\text{Tr}[\lambda^0]}\partial^\mu\theta (\fpi\tfpi^T)^{0b}\tilde a^b_\mu
\right),
\label{modI1}
\end{eqnarray}
and
\begin{equation}
\delta {\cal I}_\Omega =
\frac{N_c}{72\pi^2}\varepsilon^{\mu\nu\rho\sigma}
\int d^4x
\left(
a^a_\mu\nabla^{ac}_\nu a^c_\rho \sqrt{6}\tilde a^0_\sigma
-\frac{1}{2}a^0_\mu\partial_\nu a^0_\rho \sqrt{6}\tilde a^0_\sigma
\right).
\label{modI2}
\end{equation}
Note that the new constant $C^a$, which corresponds to 
$C$ in the SU(2) case,  now has a flavor index.
\end{enumerate}
With these modifications, the new current and density operators
$\hat{\cal O} = (\vectorjc{a}{\mu},\axialjc{a}{\mu},
\hat{\sigma}\sp{P},\hat{\pi}\sp{P},\hat{\omega})$
are defined by
$\hat{\cal O}=-i\smatrix^\dag\delta \smatrix /\delta\phi$ with
$\phi = (v^a_\mu,a^a_\mu,Y^P,J^P,-\theta)$.
The relations between the new and original operators are given by
\begin{equation}
V\sp{a}_\mu  =  
  \vectorjc{a}{\mu} 
+ \ubexav{ab}{\mu} (\fpi \hat{\pi})^b
+ \diff{(\delta {\cal I}_\Omega)}{v^{\mu a}},
\label{vectorjc}
\end{equation}
\begin{equation}
A\sp{a}_\mu      =  
  \axialjc{a}{\mu}
- \nabla^{ab}_\mu(\tfpi\hat{\pi})^b
+ (\tfpi\tfpi^T)^{ab} \tilde a_\mu^b
- (\tfpi \fpi^T)^{a0}\frac{1}{\text{Tr}[\lambda^0]}\partial_\mu\theta
+ \diff{(\delta {\cal I}_\Omega)}{a^{\mu a}},
\label{axialjc}
\end{equation}
\begin{equation}
\Sigma^a  =  G^{aP}\hat\sigma^P + C^a,
\label{sigma}
\end{equation}
\begin{equation}
\Pi^a     =  G^{aP}\hat\pi^P,
\label{pi}
\end{equation}
\begin{eqnarray}
\omega      &=&  
  \hat\omega
+ \frac{1}{\text{Tr}[\lambda^0]}\{A^{0a}[(f_\pi^{-1})^T]^{aP} 
- (\delta f_\pi)^{0P} \Box\}\hat\pi^P
\nonumber\\
&&
+ \frac{1}{\text{Tr}[\lambda^0]}(\delta f_\pi f_\pi^T)^{0a}\partial^\mu a^a_\mu
+ \frac{1}{(\text{Tr}[\lambda^0])^2}
  [ A^{00} + (\delta f_\pi \delta f_\pi^T)^{00}\Box ] \theta
+ \diff{(\delta {\cal I}_\Omega)}{(-\theta)}.
\label{omegahat}
\end{eqnarray}
As in the SU(2) case, we can identify 
the new pseudoscalar density $\hat\pi^P$ with
the (normalized) interpolating pion field
satisfying $\bra{0}\hat\pi^A(x)\ket{P(k)}=\delta^{AP}e^{-ikx}$.

The modifications induced by Eqs.~(\ref{J}), (\ref{Y}), and~(\ref{modI1}) 
can be understood as a straightforward extension of, respectively, 
Eqs.~(\ref{su2-mod11}), (\ref{su2-mod12}), and~(\ref{su2-mod22})
in the SU(2) case.
With the modification~(\ref{J}), the one-pion component is separated
from $A_\mu^a$ and $\omega$, and the resulting new operators
$\axialjc{a}{\mu}$ and $\hat\omega$
have no one-pion component surviving on pion mass-shell,
i.e., $\bra{0}\axialjc{a}{\mu}\ket{P(p)} = \bra{0}\hat\omega\ket{P(p)}= 0 $ at $p^2=(\mpi^2)^{PP}$.
From Eq.~(\ref{sigma}), the new constant $C^a$
again carries part of the quark-antiquark condensate:
$\vev{\Sigma^a}=G^{aP}\vev{\sigma^P}+C^a$.
The terms in Eq.~(\ref{modI1}), except for the first one,
are introduced such that the two-point functions including 
$A_\mu^a$ and/or $\omega$ have the correct chiral limit
in the presence of the pions.

A comment on the term $\delta{\cal I}_\Omega$ is needed
since it does not appear in the SU(2) case.
As discussed by Kaiser and Leutwyler~\cite{Kai00}, 
the non-Abelian anomaly, $\Omega^a$, which includes the 
external singlet axial gauge field $a^0_\mu$, is not invariant 
under a change of the QCD scale because of the U$_A$(1) anomaly.
This leads to an inconsistency with the RG invariance
of the VB equations\footnote{The RG invariance of the VB equations
arises from that of the S-matrix ${\smatrix}$.}.
The term $\delta{\cal I}_\Omega$ 
is introduced to cure the inconsistency and 
corresponds to the sum of the two contact terms,
$P_1$ and $P_2$, of Eq.~(78) in Ref.~\cite{Kai00}.
With this additional term the non-Abelian anomaly $\Omega^a$
is replaced with the RG invariant $\Omega^a_0$,
which is of the same form as $\Omega^a$, but
$a^0_\mu$ is replaced with $-(1/\text{Tr}[\lambda^0])\partial_\mu\theta$.

Taking account of the modifications described above,
the VB equations~(\ref{u3vvb}) and~(\ref{u3avb}) can be rewritten as
\begin{equation}
\nabla^{\mu ab}\vectorjc{b}{\mu}
+\ubexav{\mu ab}{}\axialjc{b}{\mu}+\uli{(G^{-1}J)}^{ab}(G\hat\pi)^b
+[\uli{(G^{-1}Y)}^{ab}+\uli{m}_q^{ab}](G\hat\sigma)^b
+\chi_{V1}^{aP}\hat\pi^P +\chi_{V2}^a=0,
\label{u3vvb3}
\end{equation}
and
\begin{eqnarray}
\left[
-\Box\delta^{PQ}-(\mpi^2)^{PQ}-(\fpi^{-1})^{Pa}K^{ab}\fpi^{bQ}
\right] \hat{\pi}^Q
&=&
\nonumber\\
&&
\!\!\!\!\!\!\!\!\!\!\!\!\!\!\!\!\!\!\!\!
\!\!\!\!\!\!\!\!\!\!\!\!\!\!\!\!\!\!\!\!
\!\!\!\!\!\!
- J^P
- (\fpi^{-1})^{Pa}(\nabla_\mu^{ab}\axialjc{\mu b}{}
  +\ubexav{ab}{\mu}\vectorjc{\mu b}{} -\text{Tr}[\lambda^a]\hat\omega)
\nonumber\\
&&
\!\!\!\!\!\!\!\!\!\!\!\!\!\!\!\!\!\!\!\!
\!\!\!\!\!\!\!\!\!\!\!\!\!\!\!\!\!\!\!\!
- (\fpi^{-1})^{Pa}\oli{(G^{-1}{\cal J})}^{ab}(G\hat\sigma)^b
+ (\fpi^{-1})^{Pa}(\Omega_0^a - \chi_A^a).
\label{u3avb3}
\end{eqnarray}
Here we have introduced
\begin{eqnarray}
\chi_{V1}^{aP} &=&
(\nabla^\mu\uli{\tilde a}_\mu)^{ab}\tfpi^{bP}
-(\uli{\tilde a}^\mu\uli{v}_\mu)^{ab}\tfpi^{bP}
-d^{abc}[(G^{-1})^T]^{bQ}
(\tfpi^T)^{Qd}(\nabla^\mu \tilde a_\mu)^d G^{cP}
\nonumber\\
&&
+d^{abc}[(G^{-1})^T]^{bQ}
 [(\fpi^T)^{Q0}\Box+(\fpi^{-1})^{Qa}A^{a0}]
  \frac{1}{\text{Tr}[\lambda^0]}\theta G^{cP},
\end{eqnarray}
\begin{eqnarray}
\chi_{V2}^a&=&
 \uli{\tilde a}^{ab}_\mu(\tfpi\tfpi^T)^{bc}\tilde a^{\mu c}
-\uli{\tilde a}^{ab}_\mu(\tfpi\fpi^T)^{b0}
 \frac{1}{\text{Tr}[\lambda^0]}\partial^\mu\theta
+\uli{(G^{-1} Y)}^{ab}C^b,
\end{eqnarray}
\begin{eqnarray}
\chi_A^a &=&
  \ubexv{ab}{\mu} (\fpi\tfpi^T)^{bc} \tilde a^{\mu c}
- (\fpi\fpi^T)^{ab}\ubexv{bc}{\mu}a^{\mu c}
- \ubexv{ab}{\mu}(\fpi \fpi^T)^{bc}
 \frac{1}{\text{Tr}[\lambda^0]}\partial^\mu\theta,
\end{eqnarray}
\begin{eqnarray}
K^{ab}&=& 
  (\nabla^\mu \nabla_\mu)^{ab} - (\ubexav{\mu}{} \ubexav{}{\mu})^{ab}
+ \oli{(G^{-1}Y)}^{ac}(G \fpi^{-1})^{cb} -\Box\delta^{ab},
\end{eqnarray}
\begin{eqnarray}
{\cal J}^P&=&
J^P - (\tfpi^T)^{Pa}(\nabla_\mu \tilde a^\mu)^a
+ [(\fpi^T)^{P0}\Box + (\fpi^{-1})^{Pa}A^{a0}]
 \frac{1}{\text{Tr}[\lambda^0]}\theta.
\end{eqnarray}
The terms $\chi_{V1}$, $\chi_{V2}$, and $\chi_{A}$ 
are attributable to the flavor symmetry breaking
and thus they vanish in the flavor symmetric limit.

In deriving Eqs.~(\ref{u3vvb3}) and~(\ref{u3avb3}),
we have made use of a relation satisfied by the new constant $C^a$,
\begin{eqnarray}
\overline{C}^{ab} &=& (Gf_\pi^T)^{ab},
\label{cfg}
\end{eqnarray}
which can be regarded as the U(3) version of Eq.~(\ref{su2-cfg}).
This relation follows from the fact that 
$\hat\pi^P$ is the normalized interpolating pion field
and can be derived by the same strategy as described 
for the SU(2) case in Appendix~\ref{app1}.
Then the mass relation (\ref{u3mass2}) can be rewritten as
\begin{eqnarray}
(\fpi\mpi^2\fpi^T)^{ab}&=&(\overline{m}_q\overline{C})^{ab}+A^{ab}.
\label{u3gor2}
\end{eqnarray}
From $(\overline{m}_q \overline{C})^T=(\overline{m}_q \overline{C})$
and $A^{ab}=0$ for $a\not = 0$ 
[recall that $A^{ab}=\text{Tr}[\lambda^a]({\cal A}\fpi^T)^b$],
one can fix $A^{ab}$ up to one constant:
$A^{ab} = \text{Tr}[\lambda^a]\text{Tr}[\lambda^b] A_{\chi}$.
Within the $1/\fpi$ expansion scheme~\cite{Yam96},
we have $\vev{\hat\sigma^P}=0$ at the leading order,
and therefore
$C^0=-(\sqrt{2/3})(\vev{\bar u u}+\vev{\bar d d}+\vev{\bar s s})$,
$C^3=-(\vev{\bar u u}-\vev{\bar d d})$, 
$C^8=-(\sqrt{1/3})(\vev{\bar u u}+\vev{\bar d d}-2\vev{\bar s s})$,
and $C^a=0$ for the other flavor indices.
In this case Eq.~(\ref{u3gor2}) reduces to 
the generalized GMOR relation given by Shore~\cite{Sho06}, which
is derived by making use of the Ward identities 
in the zero-momentum (soft pion) limit.
Because the leading order of the $1/\fpi$ expansion scheme
corresponds to taking the soft pion limit,
we can identify the constant $A_{\chi}$ with the nonperturbative
coefficient appearing in the gluon topological susceptibility 
in QCD~\cite{Sho06}.
The quantity $\vev{\hat\sigma^P}$ again represents the deviation of 
the mass relation (\ref{u3gor2}) from the GMOR relation 
in the same way as for the SU(2) case.

The functional derivative form of Eqs.~(\ref{u3vvb3}) and~(\ref{u3avb3})
is given by
\begin{eqnarray}
T_V^a(x)\smatrix&=&0,
\label{u3vmf}
\\
T_A^a(x)\smatrix&=&0,
\label{u3amf}
\end{eqnarray}
with
\begin{eqnarray}
T_V^a(x)&=& 
 \nabla^{ab}_\mu \diff{}{v^b_\mu(x)} +\ubexav{ab}{\mu}(x)\diff{}{a^b_\mu(x)}
+ \uli{(G^{-1}J)}^{ab}(x)G^{bP}\diff{}{J^P(x)} 
\nonumber\\
&&
+[\uli{(G^{-1}Y)}^{ab}(x)+\uli{m}^{ab}_q]G^{bP}\diff{}{Y^P(x)}
+\chi_{V1}^{aP}(x)\diff{}{J^P(x)} +i\chi_{V2}^a(x),
\label{tv}
\end{eqnarray}
\begin{eqnarray}
T_A^a(x) &=&
\fpi^{aP} 
\left[-\Box\delta^{PQ} - (m_\pi^2)^{PQ}\right]\diff{}{J^Q(x)}
+\fpi^{aP} R^P(x),
\label{ta}
\end{eqnarray}
\begin{eqnarray}
R^P(x)&=&
- (f_\pi^{-1})^{Pa} K^{ab}(x) \fpi^{bQ} \diff{}{J^Q(x)}
+ iJ^P (x)
\nonumber\\
&&
+ (f_\pi^{-1})^{Pa} 
\left[
 \nabla^{ab}_\mu \diff{}{a^b_\mu(x)} 
+\ubexav{ab}{\mu}\diff{}{v^b_\mu(x)} +\text{Tr}[\lambda^a]\diff{}{\theta(x)}
\right]
\nonumber\\
&&
+(\fpi^{-1})^{Pa}\oli{(G^{-1}{\cal J})}^{ab}(x)G^{bQ} \diff{}{Y^Q(x)}
- i(\fpi^{-1})^{Pa}[\Omega_0^a(x)-\chi_A^a(x)],
\label{u3r}
\end{eqnarray}

By introducing retarded and advanced Green's functions satisfying
\begin{eqnarray}
\left[-\Box_x\delta^{PQ} - (m_\pi^2)^{PQ} - 
(f_\pi^{-1})^{Pa} K^{ab}(x) \fpi^{bQ}\right]G^{QR}_{R,A}(x,y)
= \delta^{PR} \delta^4(x-y),
\nonumber
\end{eqnarray}
one can formally solve Eq.~(\ref{u3amf}) as
\begin{eqnarray}
\diff{}{J^P(x)}\smatrix &=&
i\smatrix\pi^P_{\text{in}} (x)
+i\smatrix\int d^4yG_R^{PQ}(x,y)
  (\fpi^{-1})^{Qa}K^{ab}(y)\fpi^{bR}\pi^R_{\text{in}} (y)
\nonumber\\
&&
-\int d^4yG_R^{PQ}(x,y)\bar R^Q(y)\smatrix
\nonumber\\
&=&
i \pi^P_{\text{in}} (x) \smatrix
+i\int d^4yG_A^{PQ}(x,y)
  (\fpi^{-1})^{Qa}K^{ab}(y)\fpi^{bR}\pi^R_{\text{in}} (y)
\smatrix
\nonumber\\
&&
-\int d^4yG_A^{PQ}(x,y)\bar R^Q(y)\smatrix,
\label{pi-solution2}
\end{eqnarray}
with
\begin{eqnarray}
\bar R^P(x) &=& 
R^P(x) + (\fpi^{-1})^{Pa} K^{ab}(x) \fpi^{bQ} \diff{}{J^Q(x)},
\end{eqnarray}
where $\pi^P_{\text{in}}$ is
the in-state asymptotic pion field.
Equations~(\ref{u3vmf}) and~(\ref{pi-solution2})
are the desired extension of the master equations
to U$_R$(3)$\times$U$_L$(3), incorporating
the finite $u,d,s$ quark masses and the U$_A$(1) anomaly.
The major consequences of flavor symmetry breaking are
the two-index character of the various constants
and the appearance of $\chi_{V1}^{aP}\pi^P$, $\chi_{V2}^a$, 
and $\chi_A^a$ terms. On the other hand, those arising from the U$_A$(1) 
anomaly are the mass relation~(\ref{u3gor2}) and
the appearance of the operator $\hat\omega$.

\subsection{Chiral reduction formula}
\label{sec3c}

The commutation relations of the pion creation and annihilation operator
with the extended S-matrix $\smatrix$ are given by
\begin{equation}
[ a^P\sb{{\rm in}}(k), \smatrix ] = R^P(k)\smatrix, 
\ \ \ \  
[ \smatrix , a^{P\dag}\sb{{\rm in}}(k) ] = R^P(-k)\smatrix,
\label{u3comm}
\end{equation}
where $R^P(k)=\int dx e^{ikx}R^P(x)$.
The U$_R$(3)$\times$U$_L$(3) version of the $\chi$RF 
for on-shell pions can be expressed as
\begin{eqnarray}
\bra{\alpha;P_1(k_1),\cdots,P_m(k_m)} 
\smatrix
\ket{\beta;Q_1(l_1),\cdots,Q_n(l_n)}|\sb{\phi=0}
&=&
\nonumber\\
&&
\!\!\!\!\!\!\!\!\!\!\!\!\!\!\!\!\!\!\!\!
\!\!\!\!\!\!\!\!\!\!\!\!\!\!\!\!\!\!\!\!
\!\!\!\!\!\!\!\!\!\!\!\!\!\!\!\!\!\!\!\!
\!\!\!\!\!\!\!\!\!\!\!\!\!\!\!\!\!\!\!\!
[ R^{P_1}(k_1)  \cdots R^{P_m} (k_m) 
  R^{Q_1}(-l_1) \cdots R^{Q_n} (-l_n) ]_{{\rm S}} 
\bra{\alpha} \smatrix \ket{\beta}|\sb{\phi=0},
\label{chrf}
\end{eqnarray}
where $k\sb{i}~(l\sb{i})$ is
the four momentum of the outgoing (incoming) pion $P_i~(Q_i)$,
and $\alpha$ and $\beta$ label states of other particles.
Here we again consider the case that no two pions have equal momenta.
The symbol $[~~]\sb{{\rm S}}$ is defined in Eq.~(\ref{sym-perm}).

Before closing this section,
we note that the singlet axial current $\axialjc{0}{\mu}$ 
(or its functional derivative form) always appears as
the RG invariant combination 
$\partial^\mu\axialjc{0}{\mu}-\text{Tr}[\lambda^0]\hat\omega$ 
in the axial master equation~(\ref{pi-solution2}) 
and the $\chi$RF~(\ref{chrf}). 
[In the absence of the external fields, 
$\axialjc{0}{\mu}$ and $\hat\omega^0$ are renormalized as
$(\axialjc{0}{\mu})_{\text{ren.}} =Z_A\axialjc{0}{\mu}$
and $ (\hat\omega)_{\text{ren.}} =
\hat\omega +(Z_A-1)(1/\text{Tr}[\lambda^0])\partial^\mu\axialjc{0}{\mu}$,
respectively.]
Therefore, the existence of $\hat\omega$, which originates from the U(1)$_A$
anomaly, is crucial for ensuring the RG
invariance of the master equations and the $\chi$RF.

\section{Applications}
\label{sec4}

As an illustration, we will present several applications
of the U$_R$(3)$\times$U$_L$(3) master equations 
and the ${\chi}$RF.

\subsection{Gluon topological susceptibility}
\label{sec4a}

By using Eqs.~(\ref{tproduct}) and~(\ref{omegahat}),
we can derive the chiral Ward identity for
the gluon topological susceptibility $\chi$:
\begin{eqnarray}
{\chi}&=&
-i\int d^4x \vev{T^\ast [\omega(x)\omega(0)]}
\nonumber\\
&=&
A_\chi -i
\int d^4x \vev{T^\ast [\hat\omega(x)\hat\omega(0)]}
\nonumber\\
&&
-i 6A_\chi^2(f_\pi^{-1})^{P0}(\fpi^{-1})^{Q0} 
\int d^4x 
\vev{T^\ast[\hat\pi^P(x) \hat\pi^Q(0)]}
\nonumber\\
&&
-i 2\sqrt{6}A_\chi(\fpi^{-1})^{P0} 
\int d^4x 
\vev{ T^\ast[\hat\omega(x) \hat\pi^P(0)]}.
\label{gsus}
\end{eqnarray}
This shows clearly how the pion poles contribute to 
the gluon topological susceptibility.
The constant $A_\chi$ of the first term corresponds to the leading contribution
in the large $N_c$ limit of QCD
with massive quarks, $\chi=A_\chi +{\cal O}(1/N_c)$~\cite{Sho07},
The appearance of $A_\chi$ is ensured by the modification introduced by
the third term of $\delta{\cal I}$ [Eq.~(\ref{modI1})].
The RG transformation property of $\hat\omega$
implies the RG invariance of 
the zero-momentum projected two-point functions
$\int d^4x \vev{T^\ast [\hat\omega(x)\hat\pi^P(0)]}$ and
$\int d^4x \vev{T^\ast [\hat\omega(x)\hat\omega(0)]}$,
and therefore Eq.~(\ref{gsus}).

Making use of the chiral Ward identities of
$\vev{T^\ast [\hat\omega(x)\hat\pi^P(0)]}$
and $\vev{T^\ast [\hat\pi^P(x)\hat\pi^Q(0)]}$ derived from 
the axial master equation~(\ref{pi-solution2})
(see Appendix~\ref{app2} for the results),
the above identity can be further written as
\begin{eqnarray}
\chi &=&
A_\chi - 6A_\chi^2[(\fpi\mpi^2\fpi^T)^{-1}]^{00}
\nonumber\\
&&
-i (1- 6A_\chi[(\fpi\mpi^2\fpi^T)^{-1}]^{00})^2
\int d^4x 
 \vev{ T^\ast[\hat\omega(x) \hat\omega(0)]}
\nonumber\\
&&
+6A_\chi^2
[(\fpi\mpi^2\fpi^T)^{-1}]^{0a}
\oli m_q^{ac} \oli{(G\vev{\hat\sigma})}^{cb}
[(\fpi\mpi^2\fpi^T)^{-1}]^{b0}.
\label{gsus2}
\end{eqnarray}
Here the second term comes from the pion pole
in $\vev{T^\ast[\hat\pi^P(x) \hat\pi^Q(0)]}$.
This is the most general expression
constrained only by the broken chiral symmetry.

We observe that our result consistently reduce to 
those obtained in previous works by taking appropriate limits.
With the mass relation~(\ref{u3gor2}), we obtain
\begin{equation}
A_\chi - 6A_\chi^2[(\fpi\mpi^2\fpi^T)^{-1}]^{00}
=A_\chi\left(1+A_\chi\sum_{q=u,d,s}\frac{1}{m_qC_q}\right)^{-1},
\label{gluonsus2}
\end{equation}
where $C_u = \tilde C+C_3$, $C_d = \tilde C -C_3$, and 
$C_s=(C_0-\sqrt{2}C_8)/\sqrt{6}$ with $\tilde C = (2C_0+C_8)/\sqrt{3}$.
The on-shell expansion scheme~\cite{Yam96}
gives $C_q=-\vev{\bar q q}$ ($q=u,d,s$) at the leading order.
In this case, the first and second terms of our general expression (\ref{gsus2})
(i.e., the contributions from the leading term in 
the large $N_c$ limit and the pion pole term)
reproduce the classic result~\cite{DiV80,Sho06}.
Also, our result approaches zero 
in the chiral limit, $\chi\to 0~(m_q\to 0)$,
showing the noncommutative character of 
the $N_c\to\infty$ and $m_q\to 0$ limits~\cite{Wit79}.

The third and forth terms in Eq.~(\ref{gsus2}) are new.
The third term shows how the higher meson states $X$,
including hybrids states and glueballs,
contribute to the gluon topological susceptibility
through $\bra{0}\hat\omega\ket{X}\not = 0$.
The forth term is proportional to $\vev{\hat\sigma^P}$, which
characterizes the deviation of the 
mass relation~(\ref{u3gor2}) from the GMOR relation.

\subsection{Two-photon decay of $\pi^0,\eta,\eta'$ mesons}
\label{sec4b}

Next let us consider the two-photon decay of 
the $\pi^0$, $\eta$, and $\eta'$ mesons: 
$P(p)\to \gamma^{(\ast)}(q_1)\gamma^{(\ast)}(q_2)$ with $P=(\pi^0,\eta,\eta')$.
For this purpose, we need to evaluate the amplitude
$\int dx_{q_1} dy_{q_2}
\bra{0}\smatrix T^\ast[\vectorjc{a}{\mu}(x) \vectorjc{b}{\nu}(y)]\ket{\pi^P(p)}|_{\phi=0}$,
where $\int dx_q\equiv \int d^4x \exp(iqx)$.
The $\chi$RF gives
\begin{eqnarray}
&&
\int dx_{q_1} dy_{q_2}
\bra{0}\smatrix T^\ast[\vectorjc{a\mu}{} (x) \vectorjc{b\nu}{} (y)]\ket{\pi^P(p)}|_{\phi=0}
\nonumber\\
&=&
\int dx_{q_1} dy_{q_2}(-i)^2\diff{}{v^a_\mu(x)}\diff{}{v^b_\nu(y)}
R^P(p)\vev{\smatrix }|_{\phi=0}
\nonumber\\
&=&
\int dx_{q_1} dy_{q_2} dz_{-p} 
(f_\pi^{-1})^{Pc}
\left[ 
i\diff{}{v^a_\mu(x)} \diff{}{v^b_\nu(y)}\Omega^c_0(z)
+i\vev{T^\ast [\vectorjc{a\mu}{}(x)\vectorjc{b\nu}{}(y)W^c(z)]}
\right],
\end{eqnarray}
where $W^c=\partial^\mu \axialjc{c}{\mu}-\text{Tr}[\lambda^c]\hat\omega$.
With expressions for the electromagnetic current, 
$j^\mu_{\text{em}} = (j_V^3)^\mu +(1/\sqrt{3})(j_V^8)^\mu$, and
the external electromagnetic field, $-eA_\mu = v^3_\mu=\sqrt{3}v^8_\mu$, 
we can derive a general expression for 
the $P\to\gamma^{(\ast)}\gamma^{(\ast)}$ 
decay amplitude:
\begin{eqnarray}
g_{\gamma\gamma P}(q_1^2,q_2^2;p^2)
&=&
\nonumber\\
&&
\!\!\!\!\!\!\!\!\!\!\!\!\!\!\!\!\!\!\!\!
\!\!\!\!\!\!\!\!\!\!\!\!\!\!\!\!\!\!\!\!
(\fpi^{-1})^{Pc}\left[c^c_{\text{em}}\frac{e^2N_c}{8\pi^2}
-\delta^{c0}\sqrt{6}F_{\gamma\gamma\hat\omega}(q_1^2,q_2^2;p^2)
+F^c_{\gamma\gamma A1}(q_1^2,q_2^2;p^2)-F^c_{\gamma\gamma A2}(q_1^2,q_2^2;p^2)
\right],
\label{Pggrel}
\end{eqnarray}
with $(c^3_{\text{em}},c^8_{\text{em}},c^0_{\text{em}})
=(2/3,2\sqrt{3}/9,4\sqrt{6}/9)$.
Here we have defined
\begin{equation}
\int dx_{q_1} dy_{q_2}
\bra{0}\smatrix T^\ast[j_{\text{em}}^\mu (x) j_{\text{em}}^\nu (y)]\ket{\pi^P(p)}|_{\phi=0}
= 
i\epsilon^{\mu\nu\alpha\beta}q_{1\alpha} q_{2\beta}
g_{\gamma\gamma P}(q_1^2,q_2^2;p^2),
\end{equation}
\begin{equation}
\int dx_{q_1} dy_{q_2} dz_{-p} 
\vev{T^\ast [j_{\text{em}}^\mu(x)j_{\text{em}}^\nu (y)\hat\omega(z)]}
= 
\epsilon^{\mu\nu\alpha\beta} q_{1\alpha} q_{2\beta}
F_{\gamma\gamma\hat\omega}(q_1^2,q_2^2;p^2),
\label{ememom}
\end{equation}
and
\begin{eqnarray}
\int dx_{q_1} dy_{q_2} dz_{-p} 
\vev{T^\ast [j_{\text{em}}^\mu(x)j_{\text{em}}^\nu (y)
\partial_\lambda\axialjc{c\lambda}{}(z)]}
&=& 
\nonumber\\
&&
\!\!\!\!\!\!\!\!\!\!\!\!\!\!\!\!\!\!\!\!
\!\!\!\!\!\!\!\!\!\!\!\!\!\!\!\!\!\!\!\!
\epsilon^{\mu\nu\alpha\beta} q_{1\alpha} q_{2\beta}
[F^c_{\gamma\gamma A1}(q_1^2,q_2^2;p^2)-F^c_{\gamma\gamma A2}(q_1^2,q_2^2;p^2)],
\label{ememja}
\end{eqnarray}
modulo $(2\pi)^4\delta^4(p-q_1-q_2)$.
In Eq.~(\ref{ememja}) we have used the 
general expression of the vector-vector-axial 
correlation function~\cite{Adl69,Sho93}
[denoting $F^c_{\gamma\gamma Ai}(q_1^2,q_2^2;p^2)$ as $F^c_{\gamma\gamma Ai}$]:
\begin{eqnarray}
\frac{1}{i}\int dx_{q_1} dy_{q_2}
\vev{T^\ast [j_{\text{em}}^\mu(x)j_{\text{em}}^\nu (y)
\axialjc{c\lambda}{}(0)]}
=
\sum_{i=1}^6 F^c_{\gamma\gamma Ai} I_i^{\lambda\mu\nu},
\nonumber
\end{eqnarray}
with
\begin{eqnarray}
&&
I_1^{\lambda\mu\nu}=\epsilon^{\lambda\mu\nu\alpha}q_{1\alpha},\ \ \ \ \ \ \ \ \,
I_2^{\lambda\mu\nu}=\epsilon^{\lambda\mu\nu\alpha}q_{2\alpha},
\nonumber\\
&&
I_3^{\lambda\mu\nu}=\epsilon^{\lambda\mu\alpha\beta}q_{1\alpha}q_{2\beta}q_2^\nu, \
I_4^{\lambda\mu\nu}=\epsilon^{\lambda\nu\alpha\beta}q_{1\alpha}q_{2\beta}q_1^\mu,
\nonumber\\
&&
I_5^{\lambda\mu\nu}=\epsilon^{\lambda\mu\alpha\beta}q_{1\alpha}q_{2\beta}q_1^\nu, \
I_6^{\lambda\mu\nu}=\epsilon^{\lambda\nu\alpha\beta}q_{1\alpha}q_{2\beta}q_2^\mu.
\nonumber
\end{eqnarray}
The broken chiral symmetry relates the $P\to\gamma^{(\ast)}\gamma^{(\ast)}$
decay amplitude to the correlation functions of the current and
density operators~(\ref{ememom}) and~(\ref{ememja}).
The correlation functions themselves, however, cannot be fixed from 
symmetry requirements and thus some dynamical inputs
are needed to evaluate them.

Equation (\ref{Pggrel}) is 
applicable both for on- and off-shell pions.
Using the PCAC hypothesis
$g_{\gamma\gamma P}(q_1^2,q_2^2;m^2_P)
\sim g_{\gamma\gamma P}(q_1^2,q_2^2;0)$ and setting
$q_1$ and $q_2$ to the photon point, $q_1^2=q_2^2=0$,
our general expression~(\ref{Pggrel}) consistently reduces to
the two-photon decay formula of $\pi^0,\eta,\eta'$ 
mesons derived by Shore~\cite{Sho06,Sho07},
\begin{equation}
g_{\gamma\gamma P}(0,0;m_P^2)\sim g_{\gamma\gamma P}(0,0;0)
=
(\fpi^{-1})^{Pc}\left[c_{\text{em}}^c\frac{e^2N_c}{8\pi^2}
-\sqrt{6}F_{\gamma\gamma\hat\omega}(0,0;0)
\right],
\end{equation}
if we identify $F_{\gamma\gamma\hat\omega}(0,0;0)$ as $Ag_{G\gamma\gamma}$.
Here we have used 
$F^c_{\gamma\gamma A1}(0,0;0)=F^c_{\gamma\gamma A2}(0,0;0)=0$~\cite{Sho93}.
The RG invariance of 
the operator $W^c=\partial^\mu \axialjc{c}{\mu}-\text{Tr}[\lambda^c]\hat\omega$
implies that the combination 
$\sqrt{6}F_{\gamma\gamma\hat\omega}
-F^0_{\gamma\gamma A1}+F^0_{\gamma\gamma A2}$ 
is also RG invariant.
Therefore $F_{\gamma\gamma\hat\omega}(0,0;0)$ is by itself RG invariant.

\section{Summary and outlook}
\label{sec5}

We have derived an extension of the master equations
for chiral symmetry breaking proposed in Ref.~\cite{Yam96,Lee99}
to the U$_R$(3)$\times$U$_L$(3) chiral group, 
carefully taking into account the U$_A$(1) anomaly and 
full flavor symmetry breaking $m_u\not=m_d\not=m_s$.
With the master equations and the $\chi$RF,
new chiral Ward identities for the gluon topological susceptibility $\chi$ and 
$P\to\gamma^{(\ast)}\gamma^{(\ast)}$ decay amplitude
have been derived, showing how the constraints from broken chiral symmetry
enter into those quantities without relying on any unphysical limits.
Then we have seen that our general results consistently reduce
to those obtained in previous studies by taking appropriate limits.

The $\chi$RF is applicable to any reaction processes
which include ground state pseudoscalar mesons, e.g.,
$\pi,K,\bar K, \eta, \eta'$ production reactions
on a baryon target and heavy meson decays such as
$J/\Psi \to 3P,\gamma 2P,\gamma 3P,\cdots$ with
$P=\pi,K,\bar K, \eta, \eta'$.
The heavy meson decays are interesting 
in relation to new meson resonance states appearing
in the decay processes~\cite{BES05,BES06}.  
A careful treatment of the final state interactions in the decay processes
will be vital for exploring properties of such new meson states.
The $\chi$RF enables one to separate details of each reaction 
mechanism from the general framework required by broken chiral symmetry,
and thus will provide a useful theoretical basis for the analysis 
of such processes.
Investigations in this direction will be discussed elsewhere.

\begin{acknowledgements}
The author would like to thank Dr. C.-H.~Lee for sending his note,
and Dr. T.~Sato and Dr. C.~Thomas for 
careful reading of the manuscript and useful comments.
This work was supported by the U.S. Department of Energy, Office of Nuclear 
Physics Division, under Contract No. DE-AC05-06OR23177
under which Jefferson Science Associates operates the Jefferson Lab.
\end{acknowledgements}

\appendix

\section{Derivation of Eq.~(\ref{su2-cfg})}
\label{app1}

In this Appendix we describe the derivation of Eq.~(\ref{su2-cfg}),
which relate the new constant $C$ to the pion decay constant
$\fpi$ and the pseudoscalar coupling constant $G$.
The strategy used here can be straightforwardly applied 
to the U(3) case [Eq.~(\ref{cfg})];
in this case the algebraic operations just become more complicated.

Originally the operator $R^a_{\text{SU(2)}}(k)$ has the following form
\begin{eqnarray}
R_{\text{SU(2)}}^a(k) &=&
\int d\sp{4}x e\sp{+ikx}
\left[
 i\frac{\fpi}{GC} J^a
+ \frac{1}{\fpi} t^a\sb{A}
- K_{\text{SU(2)}}^{ab}\diff{}{J^b} 
- \left( \nabla^{\mu ac}a_\mu^{c} - \frac{J^a}{\fpi} \right) \diff{}{Y}
\right.
\nonumber\\
&&
\left.
\ \ \ \ \ \ \ \ \ \ \ \ \ \ \ \ 
+i\left(1-\frac{\fpi}{GC}\right) \fpi\nabla^{\mu ab}a_\mu^b
\right](x).
\label{app-a1}
\end{eqnarray}
Equation~(\ref{su2-cfg}) is obtained 
by making use of the fact that the $\hat\pi^a$
is the {\it normalized} interpolating pion field satisfying
$\bra{0}\hat\pi^a(x)\ket{\pi^b(p)}=\delta^{ab}e^{-ipx}$.
By applying the $\chi$RF to the matrix element 
$\bra{0}\hat\pi^a(x)\ket{\pi^b(p)}$, we have (note that 
$\smatrix\ket{0}=\smatrix^\dag\ket{0}=\ket{0}$
following from the stability of the vacuum state),
\begin{eqnarray}
\delta^{ab}e^{-ipx}=
\bra{0}\hat\pi^a(x)\ket{\pi^b(p)}
=
-i\bra{0}\smatrix^\dag\diff{}{J^a(x)}R^b_{\text{SU(2)}}(-p)\smatrix\ket{0}|_{\phi=0},
\label{app-a2}
\end{eqnarray}
with
\begin{eqnarray}
-i\smatrix^\dag\diff{}{J^a(x)}R^b_{\text{SU(2)}}(-p)\smatrix &=&
+\frac{\fpi}{GC}\delta^{ab}e^{-ipx}
+\frac{\hat\sigma(x)}{\fpi}\delta^{ab}e^{-ipx}
\nonumber\\
&&
+\frac{i}{\fpi}\int d^4 y e^{-ipy}(ip)^\mu
T^\ast[\hat\pi^a(x)\axialjc{b}{\mu}(y)]
+ {\cal O}(\phi)
\nonumber\\
&=&
+\frac{\fpi}{GC}\delta^{ab}e^{ipx}
+\frac{\delta^{ab}}{\fpi}(p^2-\mpi^2)\int d^4 y e^{-ipy}
\Delta_R(x-y)\hat\sigma(y)
\nonumber\\
&&
-\frac{1}{\fpi}\int d^4 y e^{-ipy}
p^\mu\axialjc{b}{\mu}(y)\pi^a_{\text{in}}(x)
\nonumber\\
&&
+\frac{1}{\fpi^2}\varepsilon^{abc}\int d^4 y e^{-ipy}
\Delta_R(x-y)[-2ip^\mu+(\partial_y)^\mu]\vectorjc{c}{\mu}(y)
\nonumber\\
&&
-\frac{1}{\fpi}\int d^4y\Delta_R(x-y)
(\partial_y)^\mu\smatrix^\dag
[\smatrix\axialjc{a}{\mu}(y),a_{\text{in}}^{b \dag}(p)]
+ {\cal O}(\phi),
\label{app-a3}
\end{eqnarray}
where $(\partial_y)^\mu = [\partial/(\partial y_\mu)]$.
In the last step of  Eq.~(\ref{app-a3}), we have used 
Eqs.~(\ref{su2js}) and~(\ref{su2rbar}) 
with $R^a_{SU(2)}$ replaced by Eq.~(\ref{app-a1}), and the relation
\begin{eqnarray}
\smatrix^\dag[\smatrix\axialjc{a}{\mu}(y),a^{b\dag}_{\text{in}}(p)]&=&
-i\diff{}{a^{\mu a}(y)}R^b_{\text{SU(2)}}(-p)\smatrix
\nonumber\\
&=&
\delta^{ab}(\partial_y)_\mu[e^{-ipy}\hat\sigma(y)]
+\varepsilon^{abc}\frac{1}{\fpi} e^{-ipy}\vectorjc{c}{\mu}(y)
\nonumber\\
&&
-\frac{1}{\fpi}p^\nu\int d^4z e^{-ipz}
T^\ast[\axialjc{a}{\mu}(y)\axialjc{b}{\nu}(z)]
+{\cal O}(\phi).
\end{eqnarray}
Noticing that $\bra{0}\axialjc{a}{\mu}\ket{\pi^b}=0$,
$\bra{0}\vectorjc{a}{\mu}\ket{0}=0$, and $p^2=\mpi^2$,
Eq.~(\ref{app-a2}) gives
\begin{equation}
1 = \frac{\fpi}{GC}.
\end{equation}

\section{Commutation relations and chiral Ward identities}
\label{app2}

\subsection{Commutation relations}

The functional derivative operators defined in Eqs.~(\ref{tv}) and~(\ref{ta})
satisfy the following commutation relations
[defining $T_V^a(k)=\int d^4x \exp(ikx)T_V^a(x)$
and $T_A^a(k)=\int d^4x \exp(ikx)T_A^a(x)$]:
\begin{equation}
[T_V^a(k),T_V^b(k')] = -f^{abc}T_V^c(k+k'),
\label{commvv}
\end{equation}
\begin{equation}
[T_V^a(k),T_A^b(k')] = -f^{abc}T_A^c(k+k'),
\label{commva}
\end{equation}
\begin{equation}
[T_A^a(k),T_A^b(k')] = -f^{abc}T_V^c(k+k').
\label{commaa}
\end{equation}
With Eqs.~(\ref{ta}) and~(\ref{commaa}) we further obtain
\begin{eqnarray}
[T_A^a(k),R^P(k')] &=& 
-(\fpi^{-1})^{Pb}f^{abc}T_V^c(k+k')
\nonumber\\
&&
+ [k'^2\delta^{PQ}-(\mpi^2)^{PQ}]\int d^4x e^{i(k+k')x}
\left[i\fpi^{aQ} + \hat d^{aQC}\diff{}{Y^C(x)}\right],
\end{eqnarray}
where $\hat d^{aBC}=d^{abc}(G^{-1})^{Bb}G^{cC}$.
If $k'_\mu$ is the on-shell pion momentum satisfying
$(k')^2=(\mpi^2)^{PP}$, we have $[T_A^a(k),R^P(k')]\smatrix = 0$.
Therefore, we can make use of the same prescription
as proposed in Ref.~\cite{Kam06-2}
to derive the off-shell extension of 
the U(3) $\chi$RF~(\ref{chrf}).

\subsection{Chiral Ward identities for the two-point functions}

Here we summarize several chiral Ward identities used for obtaining
the results in Sec.~\ref{sec4a}.

Making use of Eq.~(\ref{omegahat}) and Eqs.~(\ref{operator}) 
and~(\ref{tproduct}),
the two-point function of $\omega$ can be expressed as
\begin{eqnarray}
\int d^4x e^{-ip(x-y)} \vev{T^\ast [\omega(x)\omega(y)]} &=&
\int d^4x e^{-ip(x-y)} \vev{T^\ast [\hat\omega(x)\hat\omega(y)]}
\nonumber\\
&+&
i\left[ 
A_\chi + \frac{1}{6}(\delta \tilde f_\pi \delta \tilde f_\pi^T)^{00}(-p^2)
\right] 
\nonumber\\
&+&
 {\cal N}^P(p^2)
\int d^4x e^{-ip(x-y)} 
\vev{T^\ast[\hat\omega(x) \hat\pi^P(y)]}
\nonumber\\
&+&
 {\cal N}^P(p^2)
\int d^4x e^{-ip(x-y)} 
\vev{T^\ast[\hat\omega(y) \hat\pi^P(x)]}
\nonumber\\
&+&
 {\cal N}^P(p^2) {\cal N}^Q(p^2)
\int d^4x e^{-ip(x-y)} 
\vev{T^\ast[\hat\pi^P(x) \hat\pi^Q(y)]},
\end{eqnarray}
with ${\cal N}^P(p^2) = 
\sqrt{6}A_\chi(\fpi^{-1})^{P0} 
+ (p^2/\sqrt{6}) (\delta \tilde \fpi)^{0P}$.
This identity reduces to that of the gluon topological 
susceptibility [Eq.~(\ref{gsus})] by setting $p_\mu=0$ and $y=0$.

Similarly, with the axial master equation~(\ref{pi-solution2}), 
the two-point function of 
the interpolating pion field can be written as
\begin{eqnarray}
\int d^4x e^{ip(x-y)}\vev{T^*[\hat\pi^A(x)\hat\pi^B(y)]} &=&
\int d^4x e^{ip(x-y)}
\vev{\hat\pi^B(y)\pi_{\text{in}}^A(x)} 
+i\int d^4x e^{ip(x-y)}\Delta_R^{AB}(x-y)
\nonumber\\
&&
+i\int d^4x e^{ip(x-y)}
\Delta_R^{AC}(x-y)(f_\pi^{-1})^{Cc}\hat d^{cBD}
\vev{\hat\sigma^{D}}
\nonumber\\
&&
- \int d^4x e^{ip(x-y)}
\int dz \Delta_R^{AC}(x-z)(\fpi^{-1})^{Cc}
\vev{T^*[W^c(z)\hat\pi^B(y)]}
\nonumber\\
&=&
i\int d^4xe^{ip(x-y)}\Delta_F^{AB}(x-y)
\nonumber\\
&+&
i\int d^4xe^{ip(x-y)}\Delta_R^{AC}(x-y)(\fpi^{-1})^{Cc}\hat d^{cBD}
\vev{\hat\sigma^{D}}
\nonumber\\
&+&
i\int d^4xe^{ip(x-y)}\Delta_R^{BC}(y-x)(\fpi^{-1})^{Cc}\hat d^{cAD}
\vev{\hat\sigma^{D}}
\nonumber\\
&+&
i\int d^4xe^{ip(x-y)}\int dz \Delta_R^{AC}(x-z)
\Delta_R^{BF}(y-z)
\nonumber\\
&&
\ \ \ \
\times 
(\fpi^{-1})^{Fd} [ (\mpi^2)^{CD} - 6A_{\chi}(\fpi^{-1})^{C0}(\fpi^{-1})^{D0}]
\hat d^{dDE}\vev{\hat\sigma^{E}}
\nonumber\\
&+&
 \int d^4xe^{ip(x-y)}\int dz dz'\Delta_R^{AC}(x-z) \Delta_R^{BD}(y-z')
\nonumber\\
&&
\ \ \ \
\times
(\fpi^{-1})^{Cc}
(\fpi^{-1})^{Dd} \vev{T^*[W^c(z)W^d(z')]}.
\label{eq:pipi}
\end{eqnarray}
Here we have used
$W^a=\partial^\mu\axialjc{a}{\mu}-\text{Tr}[\lambda^a]\hat\omega$;
$\Delta^{AB}_R(x-y)$ and $\Delta^{AB}_F(x-y)$ are respectively the retarded 
and Feynman propagators satisfying 
$[-\Box_x\delta^{PA} - (\mpi^2)^{PA}]\Delta^{AB}_{R(F)}(x-y)=
\delta^{PB}\delta^4(x-y)$.
In the last step, we have made use of Eq.~(\ref{pi-solution2}) and
the following Ward identities:
\begin{eqnarray}
\int d^4x e^{ip(x-y)}
\vev{T^*[\axialjc{a}{\mu}(x)\hat\pi^A(y)]}
&=&
\int d^4x e^{ip(x-y)}
\vev{\axialjc{a}{\mu}(x)\pi_{\text{in}}^A(y)}
\nonumber\\
&+&
\int d^4x e^{ip(x-y)}
p_\mu
\Delta_R^{AC}(y-x)(\fpi^{-1})^{Cc}(\tilde \fpi^{T})^{Ba}
\hat d^{cBD}\vev{\hat\sigma^{D}}
\nonumber\\
&-&
\int d^4x e^{ip(x-y)}
\int dz \Delta_R^{AC}(y-z)(f_\pi^{-1})^{Cc}
\vev{T^*[\axialjc{a}{\mu}(x)W^c(z)]},
\nonumber\\
\end{eqnarray}
\begin{eqnarray}
\int d^4x e^{ip(x-y)} \vev{T^*[\hat\omega(x)\hat\pi^A(y)]}
&=&
\int d^4x e^{ip(x-y)} \vev{\hat\omega(x)\pi_{\text{in}}^A(y)}
\nonumber\\
&-&
i \int d^4x e^{ip(x-y)} 
\sqrt{6}A_\chi\Delta_R^{AC}(y-x)(\fpi^{-1})^{Cc}(\fpi^{-1})^{B0}
\hat d^{cBD}\vev{\hat\sigma^{D}}
\nonumber\\
&-&
i \int d^4x e^{ip(x-y)} \frac{p^2}{\sqrt{6}}
\Delta_R^{AC}(y-x)(\fpi^{-1})^{Cc}[\delta \fpi^{T}]^{Ba}
\hat d^{cBD}\vev{\hat\sigma^{D}}
\nonumber\\
&-&
\int d^4x e^{ip(x-y)}\int dz \Delta_R^{AC}(y-z)(\fpi^{-1})^{Cc}
\vev{T^*[\hat\omega(x)W^c(z)]}.
\nonumber\\
\label{eq:wpi}
\end{eqnarray}

Setting $p^2=0$ and $y=0$, Eqs.~(\ref{eq:pipi}) and~(\ref{eq:wpi}) become
\begin{eqnarray}
(\fpi^{-1})^{A0} (\fpi^{-1})^{B0}
\int d^4x \vev{T^*[\hat\pi^A(x)\hat\pi^B(0)]} &=&
-i [(\fpi\mpi^2\fpi^T)^{-1}]^{00}
\nonumber\\
&&
-i2[(\fpi\mpi^2\fpi^T)^{-1}]^{0c}(\fpi^{-1})^{C0}\hat d^{cCD}\vev{\hat\sigma^D}
\nonumber\\
&&
+i [(\fpi\mpi^2\fpi^T)^{-1}]^{0a}[(\fpi\mpi^2\fpi^T)^{-1}]^{0c}
\nonumber\\
&&
\ \ \ \ \ \
\times
[(\fpi\mpi^2\fpi^T)^{ab}-A^{ab}](\fpi^{-1})^{Cb}\hat d^{cCD}\vev{\hat\sigma^D}
\nonumber\\
&&
+6 [(\fpi\mpi^2\fpi^T)^{-1}]^{00} [(\fpi\mpi^2\fpi^T)^{-1}]^{00}
\nonumber\\
&&
\ \ \ \ \ \
\times
\int d^4x\vev{T^\ast[\hat\omega(x)\hat\omega(0)]},
\end{eqnarray}
and
\begin{eqnarray}
(\fpi^{-1})^{A0}
\int d^4x \vev{T^*[\hat\omega(x)\hat\pi^A(0)]}&=&
i\sqrt{6}A_\chi
[(\fpi\mpi^2\fpi^T)^{-1}]^{0c}(\fpi^{-1})^{B0}\hat d^{cBD}\vev{\hat\sigma^D}
\nonumber\\
&&
- \sqrt{6}
[(\fpi\mpi^2\fpi^T)^{-1}]^{00}\int d^4x\vev{T^\ast[\hat\omega(x)\hat\omega(0)]},
\end{eqnarray}
respectively.
Substituting these equations into Eq.~(\ref{gsus}), we obtain Eq.~(\ref{gsus2}).


\end{document}